\documentclass[aps,prd,showpacs,nofootinbib,twocolumn]{revtex4-1}
\usepackage{natbib} 
\usepackage{amsmath, amssymb, amsbsy, latexsym, amsfonts, amsthm, bm}
\usepackage[mathscr]{eucal}
\usepackage{mathrsfs}
\usepackage{graphicx,calc,epsfig}

\newcommand{\Regge}{\mathrm{R}}
\newcommand{\R}{\mathbb{R}}
\newcommand{\C}{\mathbb{C}}
\newcommand{\so}{\mathfrak{so}}

\newcommand{\dual}{\star}
\newcommand{\sgn}{\mathrm{sgn}}

\newcommand{\BI}{\gamma}

\newcommand{\dummy}{\rule{0in}{0in}}

\begin{document}

\title{A spin-foam vertex amplitude with the correct semiclassical limit}

\author{Jonathan Engle}
\affiliation{Department of Physics, Florida Atlantic University, 777 Glades Road, Boca Raton, FL 33433, USA}

\begin{abstract}

Spin-foam models are hoped to provide a dynamics for loop quantum gravity. These start from the Plebanski formulation of gravity, in which gravity is obtained from a topological field theory, BF theory, through constraints, which, however, select 
more than one gravitational sector, as well as an unphysical degenerate sector.
We show this is why terms beyond the needed Feynman-prescribed one appear in the semiclassical limit of the EPRL spin-foam amplitude.  By quantum mechanically isolating a single gravitational sector, we modify this amplitude, yielding a spin-foam amplitude for loop quantum gravity with the correct semiclassical limit.

\end{abstract}


\maketitle

\newcommand{\mysect}[1]{\paragraph*{#1}} 

\mysect{Introduction}

Loop quantum gravity (LQG) \cite{rovelli2004, al2004, thiemann2007, rovelli2011}
offers a compelling kinematical framework in which discreteness of geometry is \textit{derived} from a quantization
of general relativity (GR) rather than postulated.  The discreteness has enabled
well-defined proposals for the Hamiltonian constraint --- defining the dynamics of the theory ---
in which one sees how diffeomorphism invariance eliminates normally problematic ultraviolet divergences.
However, the lack of manifest space-time covariance, inherent in any canonical approach to gravity,
is often suspected as a reason for the presence of ambiguities in the quantization of the Hamiltonian constraint.
This has motivated the \textit{spin-foam} program
\cite{reisenberger1994, rr1996, perez2003, rovelli2004},
which aims to provide a space-time covariant, path integral version of the dynamics of LQG a la Feynman.
The histories summed over in the path integral arise from loop quantization methods, each representing a `quantum space-time',
and referred to as a \textit{spin-foam}.

At the heart of the path integral approach  is the prescription that the contribution to
the transition amplitude by each history should be the exponential of $i$ times the action. The use of such an
expression has roots tracing back to Paul Dirac's \textit{Principles of Quantum Mechanics} \cite{dirac1930},
and is central to the successful
derivation of the classical limit of the path integral. 
In spin-foams, the `quantum space-times' have a classical geometric interpretation only in the semiclassical limit $\hbar \rightarrow 0$.
It is in this limit that one seeks a spin-foam amplitude equal to the exponential of $i$ times the classical action.
We call this the `semiclassical limit' of a spin-foam amplitude,
following \cite{cf2008}.
As highlighted in these remarks, having
such a correct semiclassical limit is key in recovering the correct
\textit{classical} limit of the theory in the standard way.

The \textit{method} used for constructing
the individual amplitudes in a spin-foam sum is to use the \textit{Plebanski} formulation of gravity, or variations thereof.  In this formulation of gravity, one takes advantage of the fact that GR can be formulated as a \textit{topological field theory} whose spin-foam quantization is well-understood --- BF theory \cite{ooguri1992} --- supplemented by so-called \textit{simplicity} constraints. Within the last several years, a spin-foam model of quantum gravity was, for the first time, introduced whose kinematics match those of LQG and therefore realize the original goal of the spin-foam program: to provide a path integral dyanamics for LQG.  This is known in the literature as EPRL \cite{epr2007, elpr2007};
when the Barbero-Immirzi parameter \cite{barbero1995, immirzi1997} $\BI$, a certain quantization ambiguity, is less than $1$, this model is equal to the Freidel-Krasnov model \cite{fk2007}. Despite its success, the EPRL amplitude still has difficulty in obtaining the correct semiclassical limit:
(non-geometric) degenerate configurations are not suppressed \cite{bdfgh2009}, and even if one restricts to non-degenerate configurations, the semiclassical limit of the simplest component of the amplitude, the vertex amplitude, has four terms instead of the desired one term of the form exponential of $i$ times the action \cite{bdfgh2009}.
Both of these problems 
cause unphysical configurations to 
dominate
in the semiclassical limit, as we will show.
(See also additional arguments \cite{oriti2004, ach2009, mv2011, clrrr2012} on the importance of having only the one 
exponential term, 
reviewed in the final discussion.)
Furthermore, we will show that both of these problems are directly due to a deficiency in the way gravity is recovered from BF theory: When one imposes the simplicity constraints, one isolates not just a single gravitational sector, but multiple sectors, not all physical.
%
%
The other 4-d spin-foam models of gravity have similar problems with a similar source \cite{bdfhp2009, bs2002, cf2008}.

In the present work, we show how, by formulating the restriction to what we call the Einstein-Hilbert sector classically first, quantizing it, and incorporating it into the EPRL vertex definition, one can define a modified vertex for which the extra terms 
in the semiclassical limit are eliminated, degenerate configurations are exponentially suppressed, and one achieves a vertex amplitude with \textit{the correct semiclassical limit}.
This new modified vertex, which we call the proper EPRL vertex, additionally continues to be compatible with loop quantum gravity, linear in the boundary state, and $SU(2)$ invariant.  The key condition of linearity in the boundary state ensures that the final transition amplitude defined by the spin-foam model is linear in the initial state and anti-linear in the final state.

To begin, we review the classical discrete framework, review the EPRL vertex, point out its problems, and then derive the solution, leading to the definition of the proper EPRL vertex.
In the final discussion, the relation of the  present work to prior work, including Teitelboim's causal propagator \cite{teitelboim1982}
and the positive frequency condition in loop quantum cosmology \cite{aps2006}, is briefly discussed.
This letter gives only a summary of this work, detailed proofs being left to the two papers
\cite{engle2011, engle2011a}.

\mysect{Review of EPRL and the problem with its asymptotics.}

The quantum histories used in spin-foam sums are usually based on a triangulation of space-time into 4-simplices.
The probability amplitude for a given spin-foam history breaks up into a product of amplitudes associated to each component of the triangulation
\cite{perez2003, rovelli2004}.
The most important of these amplitudes is the \textit{vertex amplitude}, which provides the probability amplitude for data associated to a single 4-simplex.
%
%
In the following, as we are concerned specifically with the vertex amplitude, for conceptual clarity, we focus on a single 4-simplex $\sigma$.
(Though the EPRL vertex has been generalized to arbitrary cells \cite{kkl2009}, we restrict ourselves
to the simplicial case, as certain key elements  will depend on the combinatorics of this case.
See final discussion.)
Let triangles and tetrahedra of $\sigma$ be denoted respectively by $f$ and $t$ and decorations thereof.
Fix a transverse orientation of each $f$ within the boundary of $\sigma$.
Furthermore, fix an affine structure, which is equivalent to fixing a flat connection $\partial_a$, on $\sigma$;
this is a pure gauge choice \cite{engle2011a}.  The basic variables for the single 4-simplex $\sigma$ consist in 5 group elements
$(G_{t} \in Spin(4))_{t \in \sigma}$,
and 20 algebra elements $(B_{tf}^{IJ}\in \so(4))_{f \in t \in \sigma}$, $I,J = 0,1,2,3$.
These are subject to constraints:
(1.) `orientation', $G_{t} \triangleright B_{ft} = - G_{t'} \triangleright B_{ft'}$, where $\triangleright$ denotes adjoint action,
(2.) `closure', $\sum_{f \in t} B_{ft} = 0$, and (3.) `linear simplicity', $(B_{ft})^{ij} = 0$, $i,j = 1,2,3$.
%
%
Each of these three constraints either restrict the allowed histories in the spin-foam sum or are imposed in the sense
that violations are exponentially suppressed.
Constraints (1.) and (2.) imply that there exists a \textit{unique two-form $B_{\mu\nu}^{IJ}$},
constant with respect to $\partial_a$, such that, for all $t,f$ with $f \in t$, \cite{bfh2009,engle2011}
\begin{equation}
\label{constB}
G_{t} \triangleright B_{ft}^{IJ} = \int_f B^{IJ} .
\end{equation}
In this letter, $\mu, \nu \dots$ denote tensor indices over $\sigma$ as a manifold.
When the constraint (3.), linear simplicity, is additionally imposed, this two-form field
$B^{IJ}_{\mu\nu}$ takes one of the three forms \cite{engle2011}
\begin{eqnarray*}
&(II\pm)& \; B^{IJ} = \pm \frac{1}{2} \epsilon^{IJ}{}_{KL} e^K \wedge e^L
\text{ for some constant }e^I_\mu \\
&(deg)& \; \epsilon_{IJKL} B^{IJ}_{\mu\nu} B^{KL}_{\rho\sigma} = 0\text{ (degenerate case)}
%
%
\end{eqnarray*}
where $\epsilon_{IJKL}$ is the Levi-Civita array, 
and the names for these sectors have been taken from \cite{bhnr2004, engle2011}.
If $B_{\mu\nu}^{IJ}$ is non-degenerate, it additionally
defines a dynamically determined orientation of $\sigma$, which we represent by its sign relative to the fixed orientation
$\mathring{\epsilon}$ of $\sigma$:
\begin{displaymath}
\omega := \sgn\left(\mathring{\epsilon}^{\mu\nu\rho\sigma} \epsilon_{IJKL} B_{\mu\nu}^{IJ} B_{\rho\sigma}^{KL}\right).
\end{displaymath}
For convenience we define $\omega = 0$ when $B^{IJ}_{\mu\nu}$ is degenerate.
Additionally, let $\nu:= \pm 1, 0$ according to whether $B^{IJ}_{\mu\nu}$ is in (II$\pm$) or (deg).
If $\nu \neq 0$, the BF Lagrangian is related to the Einstein-Hilbert Lagrangian by
\begin{displaymath}
\mathcal{L}_{BF} = \omega \nu \mathcal{L}_{EH}.
\end{displaymath}
When $\omega \nu = +1$, $\mathcal{L}_{BF} = \mathcal{L}_{EH}$ and we say that $B_{\mu\nu}^{IJ}$,
and the data $(B_{ft}^{IJ}, G_t)$ determining $B_{\mu\nu}^{IJ}$, are in the \textit{Einstein-Hilbert sector}.

The boundary phase space giving rise to
the $Spin(4)$ BF Hilbert space associated to the boundary of $\sigma$ is parameterized by the five group elements
$(G_{t_f't_f} \in Spin(4))_{f \in \sigma}$ and the algebra elements $(B_{ft}^{IJ})$,
where  $t_f, t_f'$ are respectively the tetrahedron `above' and `below' $f$ within the boundary of $\sigma$, and
$G_{t't} := G_{t'}^{-1} G_{t}$.
The Poisson bracket
structure is such that the combination
%
%
\begin{displaymath}
J^{IJ}_{ft} := \frac{1}{8\pi G} \left(B_{ft} + \frac{1}{\BI} \dual B_{ft}\right){}^{IJ}
\end{displaymath}
generates left or right translations on $G_{t_f't_f}$ according to whether
$t$ equals $t_f'$ or $t_f$, respectively.  The corresponding generators of (internal) spatial rotations
are then $L^i_{ft} := \frac{1}{2} \epsilon^i{}_{jk} J^{jk}_{ft}$.

In quantum theory, the simplicity constraint reduces the boundary Hilbert space of the quantum BF theory
to precisely that of LQG, yielding an embedding of LQG boundary states into $Spin(4)$ BF theory boundary states \cite{elpr2007}.
Let us recall this embedding both because it is at the heart of the EPRL vertex amplitude, and because it will be key
in the modification we propose.

%
%
The LQG Hilbert space associated to $\partial \sigma$ is $L^2(\times_f SU(2))$.
A (generalized) \textit{spin-network} $\Psi_{(k_f, \psi_{ft})}$ in this space is labeled by
one spin $k_f$ and two states $\psi_{ft_f'} \in V_f^*$, $\psi_{ft_f} \in V_f$ per triangle $f$,
%
%
where $V_{k}$ denotes the spin-$k$ representation of $SU(2)$. $\Psi_{(k_f, \psi_{ft})}\in L^2(\times_f SU(2))$
is given explicitly by
\begin{equation}
\label{snstate}
\Psi_{(k_f, \psi_{ft})}((g_f)) := \prod_{f}
\langle \psi_{ft_f'} \rho_{k_f}(g_f) \psi_{ft_f} \rangle
\end{equation}
where $\rho_k(g)$ denotes the representation matrix for $g \in SU(2)$ on $V_k$.
The embedding
$\iota: L^2(\times_f SU(2)) \rightarrow L^2(\times_f Spin(4))$ from LQG states to $Spin(4)$ BF theory boundary states
is defined in terms of the
basis (\ref{snstate}) by
\begin{displaymath}
\left(\iota \Psi_{(k\!{}_f, \psi\!{}_{ft}\!)}\right)\!\!((\!G_f\!))\!\!
:= \!\!\prod_f \! \langle \psi_{ft'\!{}_f} | \iota^{k\!{}_f} \! \rho_{s_f^-\!, s_f^+}(G_f) \iota_{k\!{}_f} | \psi_{ft\!{}_f}\! \rangle
\end{displaymath}
where here and throughout this letter we set \mbox{$s^\pm := \frac{1}{2}|1 \pm \BI| k$}, \,
$\rho_{s^-, s^+}(G)$ denotes the spin $(s^-, s^+)$ representation of $G \in Spin(4)$,
$\iota_{k} : V_k \rightarrow V_{s^-} \otimes V_{s^+}$ denotes the indicated intertwiner scaled such
that it is isometric in the Hilbert space inner products, and $\iota^k : V_{j^-} \otimes V_{j^+} \rightarrow V_k$ is its
Hermitian conjugate.

In terms of the family of states (\ref{snstate}), the operator $\hat{L}^i_{ft}$, the quantization of $L^i_{ft}$ in the $Spin(4)$ quantum theory, acts directly
on $\psi_{ft}$ as can be checked:
%
%
\begin{displaymath}
\hat{L}^i_{ft} \iota \Psi_{(k_{f'}, \psi_{f't'})} = \iota \Psi_{(k_{f'}, \tilde{\psi}_{f't'})}
\end{displaymath}
where $\tilde{\psi}_{ft} := \hat{L}^i \psi_{ft}$, and $\tilde{\psi}_{f't'}:= \psi_{f't'}$ for $f' \neq f$ or $t \neq t'$,
and where $\hat{L}^i$ denotes the $SU(2)$ generators acting in the appropriate irreducible representation.
%
%

The \textit{EPRL vertex amplitude} $A_\sigma^{EPRL}: L^2(\times_f SU(2)) \rightarrow \C$, in terms of the
above is
\begin{eqnarray}
\nonumber
&& \dummy \!\!\!\!\!\! A_\sigma^{EPRL}(\Psi_{(k_f, \psi_{ft})})
\! := \!\!\!\!\!\!
\underset{Spin(4)^5}{\int} \!\!\! \Big(\! \prod_t \! d G_t \!\Big) \left(\iota \Psi_{(k_f, \psi_{ft})}\right)((G_{t_f't_f}\!)) \\
\label{lqgvert}
&& \dummy \!\!
=\! \int_{Spin(4)^5} \! \Big(\! \prod_t \! d G_t \!\Big) \!\! \prod_f \!
\langle \psi_{ft_f'} | \iota^{k_f} \rho(G_{t_f' t_f}\!) \iota_{k_f}
|\psi_{ft_f} \! \rangle .
\end{eqnarray}
This vertex amplitude can be rewritten using
\textit{Livine-Speziale} coherent states \cite{ls2007}. Each such state $\Psi_{(k_f, n_{ft})}$ is labeled by one spin $k_f$ per $f$, and one unit 3-vector $n_{ft}$ per $f,t$ with $f \in t$, and are obtained from the states (\ref{snstate}) by setting
$\psi_{ft_f'} := \langle k_f, -n_{ft_f'} |$ and $\psi_{ft_f} := | k_f, n_{ft_f}\rangle$,
where $|k, n \rangle \in V_k$ denotes the $SU(2)$ \textit{Perelomov} coherent state \cite{perelomov1972} 
satisfying $\langle k, n | \hat{L}^i | k,n \rangle = k n^i$.
One then has
\begin{eqnarray}
\label{cohlqgvert}
&& \dummy \!\!\!\!\!\! A_\sigma^{EPRL}(\Psi_{(k_f, n_{ft})}) = \\
\nonumber
&& \dummy \!\!\!\!\!\!
\int_{Spin(4)^5} \! \Big(\! \prod_t \! d G_t \!\Big) \!\! \prod_f \!
\langle k_f, \! n_{ft_f'} | \iota^{k_f} \rho(G_{t_f' t_f}\!) \iota_{k_f}
|k_f, \! n_{ft_f} \! \rangle
\end{eqnarray}
%
%
The $Spin(4)$ boundary state $\iota \Psi_{(k_f, n_{ft})}$ is a coherent state peaked on the classical configuration of
$B_{ft}^{IJ}$'s taking the values
\begin{equation}
\label{Bvals}
B_{ft}^{IJ} = 16 \pi G k_f \delta^{[I}_{0} n^{J]}_{ft}
\end{equation}
where $n^0_{ft} := 0$.  The $B_{ft}^{IJ}$ values of the form (\ref{Bvals}) are the most general satisfying
linear simplicity (constraint (3.) enumerated earlier). Furthermore, the integral over the $G_t$'s can
be interpreted as a \textit{path integral} over possible $G_t$'s, which one identifies with the parallel transports
introduced in the discrete classical framework reviewed above.
This identification and interpretation of the group integration variables in (\ref{cohlqgvert}) is justified by the
large spin ($k_f$) asymptotic analysis of (\ref{cohlqgvert}) carried out by Barrett et al.~\cite{bdfgh2009}, in which one finds that the critical point equations for the $G_t$'s are precisely those satisfied by the parallel transports in the discrete classical framework.

If the data $(k_f, n_{ft})$ is such that,
for each $t$, the span of $\{n_{ft}\}_{f \in t}$ is three dimensional,
and is such that there exist group elements $G_t$ allowing all the constraints (1.), (2.), and (3.) to be
satisfied, then a unique \textit{Regge geometry} \cite{regge1961} of the 4-simplex is determined and the data
are called \textit{Regge-like}.
%
%
In this case
the overall phase of the coherent state $\Psi_{(k_f, n_{ft})}$
can be fixed uniquely, giving rise to what is called the \textit{Regge state} $\Psi_{(k_f, n_{ft})}^\Regge$
\cite{bdfgh2009}.
For such states, the asymptotics of the EPRL vertex are
\begin{eqnarray}
\label{lqgasym}
\dummy \hspace{-0.1cm} 
A_\sigma^{EPRL}(\Psi^\Regge_{(\lambda k_f, n_{ft})}) &\sim& \lambda^{-12} \big(N_1 e^{iS_\Regge} + N_1 e^{-iS_\Regge} \\
\nonumber
&& \qquad + N_2 e^{\frac{i}{\BI}S_\Regge} + N_3 e^{-\frac{i}{\BI}S_\Regge}\big) ,
\end{eqnarray}
where $\sim$ indicates that the error term is bounded by a constant times the power of $\lambda \in \R^+$ indicated, $S_\Regge$ denotes the Regge action determined by the data $(\lambda k_f, n_{ft})$,
and $N_i$ are real functions of $(k_f, n_{ft})$.
%
%

One sees that the presence of the four distinct terms in (\ref{lqgasym})
spoils the classical limit of the model when multiple 4-simplices are involved.
Consider a spin-foam on a triangulation $\Delta$ whose data we assume, for simplicity, 
is Regge-like at each 4-simplex.
The full amplitude then takes the form $A(\Delta) = \prod_{f} A_f \prod_t A_t \prod_\sigma A_\sigma^{EPRL}$,
where $A_f$ and $A_t$ are the factors associated to each triangle $f$ and tetrahedron $t$ in $\Delta$ \cite{elpr2007}.
Let $S_{f,t}$ denote the phase angle of the product
$\prod_{f} A_f \prod_t A_t$ (defined modulo $2\pi$);
%
%
as $A_f$ and $A_t$ are always real, $S_{f,t}$ is $0$ or $\pi$.
The asymptotics of \textit{each} factor $A_\sigma^{EPRL}$ now has four terms as in (\ref{lqgasym}).
On multiplying out these terms, the semiclassical limit of the full amplitude takes the form
\begin{displaymath}
\textstyle A(\Delta) \sim \sum_{(\lambda_\sigma \in \{\pm 1, \pm 1/\BI\})} N_{(\lambda_\sigma)} e^{iS_{(\lambda_\sigma)}},
\end{displaymath}
where $S_{(\lambda_\sigma)}:= S_{f,t} + \sum_\sigma \lambda_\sigma S_R(\sigma)$,
and the sum is over all possible ways of choosing the coefficient $\lambda_\sigma$
for each simplex $\sigma$ to be the coefficient of the action in one of the four terms in (\ref{lqgasym}).
$S_{(\lambda_{\sigma})}$ is the Regge action for $\Delta$ (modulo $2\pi$)
if and only if all the $\lambda_\sigma$ are $1$.
Because, however, the $\lambda_\sigma$ can vary from 4-simplex to 4-simplex,
$S_{(\lambda_{\sigma})}$ is in general \textit{not} equal to the Regge action, even upto to rescaling by a constant, and
its stationary points do not in general solve the Regge equations of motion.
One thus has sectors in the semiclassical limit of the model which do not represent general relativity.
This is in addition to the spin-foams which persist in the semiclassical limit whose data are degenerate and do not
even represent a space-time geometry.

The most obvious way to correct the problem with the semiclassical limit of the amplitude for Regge-like data
is to somehow alter the vertex amplitude such that all but the first term in (\ref{lqgasym}) is eliminated.  
How might one do this?
Each term in the asymptotics (\ref{lqgasym}) corresponds to a critical point of the integral (\ref{cohlqgvert}),
and hence to a particular value of the variables $G_t$, which, together with the boundary data (\ref{Bvals}),
as mentioned earlier, determine a continuum two-form $B_{\mu\nu}^{IJ}$ which is in one of the three Plebanski sectors
labelled by $\nu = 0, \pm 1$, and which determines an orientation labelled by
$\omega= 0, \pm 1$. The values of $\nu$ and $\omega$ corresponding to each
of the four terms in (\ref{lqgasym}) satisfy $\omega\nu = +1, -1, 0, 0$, respectively.
Therefore, to isolate the first term, one must impose $\omega\nu = +1$  --- that is,
one must restrict to the \textit{Einstein-Hilbert sector} as we have defined it.  
This, at the same time, will eliminate the degenerate sector.

\mysect{A condition selecting the Einstein-Hilbert sector and its quantization.}

Our strategy is first to find a classical condition on the basic variables that selects the
Einstein-Hilbert sector,
quantize this condition, and then use it to modify the EPRL vertex (\ref{lqgvert}, \ref{cohlqgvert}).
For each two tetrahedra $t,t'$ define the sign $\beta_{tt'}((G_{\tilde{t} \tilde{t}'}))$ by
\begin{eqnarray*}
\beta_{tt'}((G_{\tilde{t} \tilde{t}'})) \! &:=& \! -\sgn\left[\epsilon_{ijk} (G_{tt_1})^i{}_0
(G_{tt_2})^j{}_0  (G_{tt_3})^k{}_0 \cdot \right.\\
&& \quad \left. \cdot \epsilon_{lmn} (G_{t't_1})^l{}_0
(G_{t't_2})^m{}_0  (G_{t't_3})^n{}_0 \right] \quad \dummy
\end{eqnarray*}
where $G^I{}_J$ denotes the $SO(4)$ matrix canonically associated to a given $Spin(4)$ element $G$
(see, e.g., \cite{bdfgh2009, engle2011a}),
$t_1, t_2, t_3$ are the tetrahedra in $\sigma$ \textit{other} than $t,t'$, in any order,
and $\sgn$ is defined to be zero when its argument is zero. We then have that
the constant 2-form $B_{\mu\nu}^{IJ}$ determined by
(\ref{constB}) is in the Einstein-Hilbert sector iff
\begin{equation}
\label{IIpluscond}
\beta_{tt'}((G_{\tilde{t} \tilde{t}'})) \cdot (G_{t})^i{}_0 \cdot (L_{ft})_i > 0
\end{equation}
for any pair $t,t'$, where $f = t \cap t'$ \cite{engle2011a}.
This is the condition which we seek to quantize and use to modify the vertex integral (\ref{lqgvert}, \ref{cohlqgvert}).
Normally this would be done by inserting into the path integral (\ref{cohlqgvert}) the Heaviside function
\begin{equation}
\label{classinsert}
\Theta\left(\beta_{tt'}((G_{\tilde{t} \tilde{t}'})) \cdot (G_{t})^i{}_0 \cdot (L_{ft})_i \right),
\end{equation}
where $\Theta(\cdot)$ is zero when its argument is zero.
However, if one inserts this into (\ref{cohlqgvert}), one obtains a vertex amplitude which is
\textit{non-linear in the boundary state}, spoiling the property of the final spin-foam sum that it be linear in the initial state and anti-linear in the final state (a property necessary for the final spin-foam sum to be interpreted as a transition amplitude).
Instead, we \textit{partially quantize} the expression (\ref{classinsert})
before inserting it into (\ref{lqgvert}, \ref{cohlqgvert}), by replacing $L_{ft}^i$ with $SU(2)$ generators $\hat{L}^i$ acting on the coherent state $|k_f, n_{ft} \rangle$.
Because the generators $\hat{L}^i$ are peaked on $L_{ft}^i = j_f n_{ft}^i$ when acting
on the coherent state $|k_f, n_{ft} \rangle$, such insertions will still impose the desired condition (\ref{IIpluscond})
in the semiclassical limit, and so remove  the unwanted sectors, while at the same time preserving the necessary linearity in the boundary state.  Thus we insert the following
group-variable dependent operator on $V_{k_f}$:
\begin{displaymath}
\hat{P}_{t't}((G_{\tilde{t}' \tilde{t}})) := P_{(0,\infty)}\left(\beta_{t' t}((G_{\tilde{t}' \tilde{t}})) \cdot (G_{t})^i{}_0 \cdot \hat{L}_i\right)
\end{displaymath}
where $P_{\mathcal{S}}(\hat{O})$ denotes the spectral projector for the operator $\hat{O}$ onto the portion $\mathcal{S}$ of its spectrum.
Inserting this into the
vertex path integral (\ref{lqgvert}), one obtains what we call the \textit{proper EPRL vertex amplitude}.
For a \textit{general} $SU(2)$ spin-network state (\ref{snstate}), it is given explicitly by
\begin{eqnarray}
\label{proplqgvert}
%
%
&& \dummy \!\!\! A_\sigma^{(+)}(\Psi_{(k_f, \psi_{ft})}) = \\
\nonumber
&& \dummy \!\!\!\!\!\!\!
\underset{Spin(4)^5}{\int} \!\!\!\!\! \Big(\! \prod_t \! d G_t \!\Big) \!\! \prod_f \!
\langle \psi_{ft{}_f'} | \iota^{k\!{}_f} \! \rho(G_{t_f' t\!{}_f}\!) \iota_{k\!{}_f}
\!\hat{P}_{t_f' t\!{}_f}((\!G_{t_f' t\!{}_f}\!))| \psi_{ft\!{}_f} \! \rangle .
\end{eqnarray}
One can equivalently write the vertex amplitude with the projector on the left side of each integrand factor \cite{engle2011a}.
This vertex amplitude is manifestly linear in the boundary state (\ref{snstate}),
%
%
and one can furthermore show that
it is $SU(2)$ invariant \cite{engle2011a}.
For the coherent state $\Psi_{(\lambda k_f, n_{ft})}$, for large $\lambda$, the proper EPRL vertex is exponentially suppressed unless
$(k_f, n_{ft})$ describes a non-degenerate Regge geometry, in which case it furthermore now has \textit{precisely the required asymptotics}
\cite{engle2011a}
\begin{displaymath}
A_\sigma^{(+)}(\Psi^\Regge_{(\lambda k_f, n_{ft})}) \sim \lambda^{-12} N_1 e^{iS_\Regge} .
\end{displaymath}

\mysect{Discussion.}

By implementing quantum mechanically a restriction to the Einstein-Hilbert sector,
the EPRL vertex amplitude has been modified, yielding
what we call the proper EPRL vertex.  The resulting vertex is linear in the boundary state, $SU(2)$ invariant, 
and leads to a correct semiclassical limit.

Let us remark first on the non-triviality of the removal of the degenerate sector that has been achieved.
In the work \cite{cf2008},
the degenerate sector of the Freidel-Krasnov model (equal to EPRL for $\BI < 1$) is removed by using a path integral representation
based on coherent states, similar to the path integral representation of the EPRL vertex
given in (\ref{cohlqgvert}) above. However, in the conclusion of the work \cite{cf2008}, the
authors mention that they do not know how to rewrite the
resulting restricted path integral as a spin-foam sum --- that is, as a sum over histories of spin-foams labeled with spins and intertwiners, similar
to (\ref{lqgvert}).
The reason for this difficulty arguably can be traced to the same reason for our rejection of the ``naive'' prescription of inserting the non-quantized Heaviside function (\ref{classinsert}) into (\ref{cohlqgvert}): because the resulting transition amplitude is non-linear in the boundary state.  Thus, as far as removal of the degenerate sector is concerned, the new element of the present work is precisely the fact that the removal is achieved in such a way that \textit{linearity in the boundary state is preserved}, so that the vertex amplitude can continue to be used to define transition amplitudes between canonical states in the usual sense.

Beyond the removal of the degenerate sector, the proper vertex furthermore achieves isolation of the Einstein-Hilbert sector, in which
the sign of the Lagrangian relative to the Einstein-Hilbert Lagrangian is restricted to be consistently positive, ensuring the correct equations of motion in the semiclassical limit.  In doing this, linearity in the boundary state is again preserved.  This contrasts with the modification proposed in \cite{mv2011}, in which the undesired term 
%
%
in the asymptotics is removed by direct means without understanding first its deeper meaning in terms of Plebanski sectors and orientations.

In addition to ensuring the correct equations of motion, the fact that the proper vertex asymptotically has only
a single term with a single sign in front of the action may solve other problems as well.
In particular, such asymptotics seem necessary in order for spin-foams to be consistent with the positive frequency
condition in loop quantum cosmology \cite{aw2009, aps2006}.  They have also been advocated by
Oriti \cite{oriti2004} as a way of implementing causality in the sense introduced by Teitelboim \cite{teitelboim1982}.
Finally, from studies of 3-d gravity, there are indications that such asymptotics
may completely eliminate a certain divergence in spin-foam sums present until now \cite{clrrr2012}.

The expression (\ref{proplqgvert}) can be easily generalized to the Lorentzian signature \cite{engle2011a}.  One open issue is to provide a derivation of this generalization as well as to verify that, like the Euclidean proper vertex above, 
it has the desired semi-classical properties. A second open issue is to generalize this work to an arbitrary cell 
\cite{kkl2009}.  This second generalization will likely require an entirely new perspective, as the combinatorics of the 4-simplex are presently used in a key way not only in the derivation of the proper vertex, but in its very definition.

\mysect{Acknowledgements.}

The author would like to thank Abhay Ashtekar, Chris Beetle, Frank Hellmann, and Carlo Rovelli for helpful exchanges.


\end{document}